\documentstyle[prl,aps,multicol,epsfig]{revtex}

\begin{document}
\large
\title{Quantum Information processing by NMR: Implementation of Inversion-on-equality gate, 
Parity gate and Fanout gate\footnote{{\small \it Dedicated to Prof. P.T. Narasimhan on his 75th birthday}}}
\author{T.Gopinath$^1$, Ranabir Das$^1$, 
 and Anil Kumar $^{1,2,}$\footnote{{\small \it DAE-BRNS Senior Scientist}\\}\\
        $^1$ {\small \it  Department of Physics, Indian Institute of Science, Bangalore, India}\\
        $^2$ {\small \it Sophisticated Instruments Facility, Indian Institute of Science, Bangalore, India}\\}

\maketitle
\vspace{2cm}
\hspace{7cm}
Abstract\\
\begin{abstract}
While quantum information processing by nuclear magnetic resonance (NMR) with small number of 
qubits is well established, implementation of 
 lengthy computations have proved to be difficult due to decoherence/relaxation. 
In such circumstances, shallow circuits (circuits using parallel computation)
may prove to be realistic. Parity and fanout gates are essential to  
create shallow circuits. In this article we implement 
inversion-on-equality gate, followed by parity gate and fanout gate in 3-qubit systems by NMR, using evolution under 
indirect exchange coupling Hamiltonian.
\end{abstract}
\pagebreak
\section{Introduction}
Quantum information processing has the promise of solving problems which are intractable by 
classical computing \cite{pw,deujoz,gr}.
 Quantum computers use the laws of quantum mechanics to operate logic gates on 
 superposition of multi particle states, thereby achieving massive parallelism \cite{gru,db,ic}.
 Although this has created a great deal of interest in 
building quantum computers, the experimental realization has been extremely difficult and success 
has been limited \cite{db,ic}. One of the main reasons for limited success has been the effect of the decoherence.
The multi particle coherent states that are used by quantum computers, are extremely vulnerable to effects 
of decoherence, thereby causing errors in lengthy computations. To counter these, researchers have developed 
quantum error correcting codes \cite{pw1} and methods for fault-tolerant quantum computation \cite{st1,st2}.
 Another approach has been the use of shallow circuits \cite{sha1,sha2,sha3,sha4}. 
Shallow circuits correspond to parallel algorithms that can be performed in small amounts of time on  
idealized parallel  computers. Several gates have proved  to be useful for
creating shallow circuits including fanout gate, parity gate and inversion-on-equality gate 
\cite{sha3,sha4}. Together with single-qubit operations, the fanout, parity and inversion-on-equality 
gates are universal for quantum computation \cite{sha4}. Fenner \cite{sha4} has 
given a protocol to implement inversion-on-equality gate, parity gate and fanout gate  using the 
evolution under a Hamiltonian of the form
\begin{eqnarray}
H_n=\sum_{1\leq i \leq j \leq n} J_{ij} Z_i Z_j
\end{eqnarray}
where $Z_i$ is the angular momentum operator of $i^{th}$ particle and $J_{ij}$ are the energy coefficients. 

Among various physical systems that have been suggested for building quantum computers \cite{db,ic}, 
nuclear magnetic resonance (NMR) has been the most successful physical system so far by 
demonstrating quantum information processing on a small number of qubits 
\cite{db,ic,na,dg,djjo,kd,ai,ci,tel,ap,fun,mulf,curr1,jo,lo,pram1,qft,ne,ts,nat,mur,ron,pram2,rd,rd1,curr2,rd2}. In 
NMR, weakly coupled spin-1/2 nuclei having sufficiently different Larmor frequencies,
 and coupled to each other by indirect exchange (J) couplings are used as qubits \cite{na,dg}.
The Hamiltonian proposed by Fenner is identical to the well known J-coupling Hamiltonian of weakly coupled 
spin-1/2 spins (qubits) in NMR, namely
\begin{eqnarray}
H_J=\sum_{i,j (i<j)} J_{ij} I_{iz}I_{jz}.
\end{eqnarray}
 
 In this work we implement inversion-on-equality gate, parity gate and fanout gate using the circuits 
proposed by Fenner, on 3-qubit systems by NMR, using evolution under the  J-coupling Hamiltonian.
To the best of our knowledge this is the first explicit experimental demonstration of these gates in 
a quantum physical system.  

\section{inversion-on-equality gate}
 Inversion-on-equality (I$_=$) gate inverts the state when the states of all the qubits are equal. 
In a 3-qubit system it act as 
\begin{eqnarray}
I_= \vert abc \rangle=(-1)^{\delta_{ab}\delta_{bc}}\vert abc \rangle. 
\end{eqnarray}
The truth table of (I$_=$) gate is given in Table 1. The unitary operator for implementation of this gate 
is of the form 
\begin{eqnarray}
U_{I_=}=\pmatrix{ -1&0&0&0&0&0&0&0 \cr 0&1&0&0&0&0&0&0 \cr 0&0&1&0&0&0&0&0 \cr 0&0&0&1&0&0&0&0 \cr
                  0&0&0&0&1&0&0&0 \cr 0&0&0&0&0&1&0&0 \cr 0&0&0&0&0&0&1&0 \cr 0&0&0&0&0&0&0&-1}
\end{eqnarray} 
The inversion-on-equality gate for a 3-qubit system can also be termed as $U_3$ (fig. 1(a)) \cite{sha4}.
Applying I$_=$ on a three qubit system, when one of the qubits is in state $\vert 1 \rangle$,
results in controlled-Z gate in the other two qubits \cite{ic}. Such gates are used  in the 
implementation of universal CNOT gates\cite{djjo}.
In a 3-qubit NMR system, this inversion gate can be implemented by evolution under J-couplings, yielding 
\begin{eqnarray}
U_{I_=}=e^{-i2\pi(J_{12} I_{1z}I_{2z}\tau_1+J_{23} I_{2z}I_{3z}\tau_2+J_{13} I_{1z}I_{3z}\tau_3)}
\end{eqnarray}
 where $\tau_1$=1/2J$_{12}$, $\tau_2$=1/2J$_{23}$ and $\tau_3$=1/2J$_{13}$. Hence this gate requires 
a $1/(2J)$ evolution for all the J-couplings. 
The three protons of  1-bromo, 2,3-dichloro benzene constitute 
a 3-qubit system. In a magnetic field of 11.4 tesla, the Larmor frequency ($\nu$) 
differences between the protons are $\vert\nu_1-\nu_2\vert$= 176 Hz, $\vert\nu_2-\nu_3\vert$=  61 Hz and 
$\vert\nu_1-\nu_3\vert$= 237 Hz.
 The J-couplings between the protons are $J_{12}$=8.1 Hz , $J_{23}$= 1.47 Hz and 
$J_{13}$=8.1 Hz. A direct but
lengthy protocol would be to use a sequence of three evolutions ($\tau_1$,$\tau_2$ and $\tau_3$),
with the required J-couplings retained during each $\tau$ and refocusing all other couplings
and chemical shifts by the well known refocusing schemes \cite{er}.

However, since $J_{12}=J_{13}$, two of evolutions can be performed in parallel, yielding a 
simpler pulse sequence. Such a pulse sequence is of the form,
\begin{eqnarray}
U_{I_=}=
\tau_1/2-(\pi)-\tau_1/2-\tau/4-(\pi)^1-\tau/4-(\pi)-\tau/4-(\pi)^1-\tau/4,
\end{eqnarray}
 where  $(\pi)^i$ is a $\pi$-pulse applied on the 
$i^{th}$ qubit and $(\pi)$ is a $\pi$-pulse applied on all the qubits (Figure 1(c)). 
 The chemical shift is refocused throughout the sequence by appropriate $\pi$-pulses. During $\tau_1/2-(\pi)-\tau_1/2$, 
 evolution takes place under all the couplings. During  
$\tau/4-(\pi)^1-\tau/4-(\pi)-\tau/4-(\pi)^1-\tau/4$, evolution takes only under $J_{23}$. 
 Choice of appropriate values for the delays such that $\tau_1=1/(2J_{12})=1/(2J_{13})$ and $\tau=(1/(2J_{23})-\tau_1)$,
 yields a $1/(2J)$ evolution for all the J-couplings, making $U_{I_=}$ of Eq. (5) equal to Eq. (4).
 In order to observe the output of $I_=$ gate operation, it is convenient to apply the gate on a 
coherent superposition of all the qubits. Such a coherent superposition can be created by 
a pseudo-Hadamard gate on the three qubits (fig. 1(b)). 
 
A pseudo-Hadamard gate is of the form \cite{djjo}
\begin{eqnarray}
h=\frac{1}{\sqrt{2}} \pmatrix{1&1\cr-1&1}.
\end{eqnarray}
Pseudo-Hadamard gate on a single qubit can be implemented by a qubit-selective $(\pi/2)_y$ pulse \cite{djjo}.
Pseudo-Hadamard gate on all three qubits $h_1\otimes h_2\otimes h_3$ can be implemented by a non-selective $(\pi/2)_y$
pulse on all the qubits. Starting from equilibrium, $h_1\otimes h_2\otimes h_3$ creates 
single quantum coherences (Figure 2(a)) and the density matrix is of the form,

\pagebreak
\begin{eqnarray}
\hspace{-2cm}
&&\matrix{\vert000\rangle&\vert001\rangle&\vert010\rangle&\vert011\rangle&\vert100\rangle
&\vert101\rangle&\vert110\rangle&\vert 111\rangle}\\ \nonumber
\sigma_1=
&&\pmatrix{ 0&~~~~~1&~~~~~1&~~~~~0&~~~~~~1&~~~~~0~~&~~~0~~~~&~~0 
   \cr 1&0&0&1&0&1&0&0 \cr 1&0&0&1&0&0&1&0 \cr 0&1&1&0&0&0&0&1 \cr
                  1&0&0&0&0&1&1&0 \cr 0&1&0&0&1&0&0&1 \cr 0&0&1&0&1&0&0&1 \cr 0&0&0&1&0&1&1&0}
\matrix{\vert000\rangle \cr \vert001\rangle \cr \vert010\rangle \cr \vert011\rangle \cr \vert100\rangle
\cr \vert101\rangle \cr \vert110\rangle \cr \vert111\rangle}
\end{eqnarray}

After this, the pulse sequence of Eq. (6) is implemented which evolves $\sigma_1$ to $\sigma_2$, 
\begin{eqnarray}
\hspace{-2cm}
&&\matrix{\vert000\rangle&\vert001\rangle&\vert010\rangle&\vert011\rangle&~~~\vert100\rangle
&\vert101\rangle&\vert110\rangle&\vert 111\rangle}\\ \nonumber
\sigma_2=U_{I_=}^{\dagger}~(\sigma_1)~U_{I_=}=
&&\pmatrix{ ~~0&~\fbox{-1}&~~\{-1\}&~~~0&~~~~\langle-1\rangle
&~~~0~~&~~~0~~~~&0 \cr
-1&0&0&\{1\}&0&\langle1\rangle&0&0
\cr -1&0&0&\fbox{1}&0&0&\langle~1\rangle&0
\cr 0&1&1&0&0&0&0&\langle-1\rangle \cr
                  -1&0&0&0&0&\fbox{1}&\{~1\}&0 \cr 0&1&0&0&1&0&0&\{-1\}
 \cr 0&0&1&0&1&0&0&\fbox{-1} \cr 0&0&0&-1&0&-1&-1&0}
\matrix{\vert000\rangle \cr \vert001\rangle \cr \vert010\rangle \cr \vert011\rangle \cr \vert100\rangle
\cr \vert101\rangle \cr \vert110\rangle \cr \vert111\rangle}
\end{eqnarray}

The four single quantum transitions of each qubit are identified with different symbols, 
transitions of 1st qubit are within angular brackets $(\langle\rangle)$, 
transitions of 2nd qubit are within curly brackets ($\{\}$), and that of 3rd qubit 
are in a box ($\fbox{}$). It may be noted that the sign of two single-quantum transitions  
of each qubit are inverted, specifically the ones 
in which the other qubits are either in $\vert 00 \rangle$ or $\vert 11 \rangle$ states.
The obtained spectrum after implementation of the inversion-on-equality gate is given in Figure 2(b).
 The expected spectrum is shown as stick diagram in Figure 2(c). The maximum error of this implementation was 5$\%$.      
 
 Fenner has suggested the use of n-qubit inversion-on-equality gates to construct (n+1)-qubit parity and fanout gates \cite{sha4}.
 In the next two sections we will implement 3-qubit parity and fanout gates by using 2-qubit inversion-on-equality gate ($U_2$).

\section{Parity gate}
 Parity gate adds (addition modulo 2) the control bits to the target bit. 
The truth table of parity gate on a 3-qubit system where third qubit is the target qubit 
and the first two qubits are control, is given in Table 2. A 3-qubit version of the 
quantum circuit proposed by Fenner \cite{sha4} is given in Figure 3(a). The circuit uses a two-qubit U$_2$ gate which 
is of the form  U$_2$=$e^{-i2\pi(J_{12} I_{1z}I_{2z}\tau)}$ where $\tau$=1/2$J_{12}$. 
U$^\dagger_2$ is of the same form but with $\tau$=3/2$J_{12}$. 
 Besides U$_2$ and pseudo-Hadamard gates, the other gates used are phase gate 
\begin{eqnarray}
s=\pmatrix{1&0\cr0&i},
\end{eqnarray}
and CNOT gate 
\begin{eqnarray}
CNOT=\pmatrix{1&0&0&0&0\cr 0&1&0&0&0\cr0&0&0&0&1\cr0&0&0&1&0}.
\end{eqnarray}
Pseudo-Hadamard gate is implemented by a $(\pi/2)_{-y}$ pulse. The phase gate requires a composite 
z-pulse of the form $(\pi/2)_{-y}(\pi/2)_{x}(\pi/2)_{y}$ \cite{djjo}. The CNOT gate between 2nd and 3rd qubit 
in the quantum circuit of Figure 3(a), can be implemented by a cascade 
[$(\pi/2)_y^3-\tau/2-(\pi)_x^{2,3}-\tau/2-(\pi/2)_x^3$] \cite{dg}. While combining the whole pulse sequence of 
the quantum circuit, some consecutive pulses of opposite phases cancel out, simplifying 
the sequence. The simplified complete sequence (fig. 3(b)) is
\begin{eqnarray} 
&&(\pi/2)_{-y}^2-\tau_1/2-(\pi)_x^{1,2}-\tau_1/2-(\pi/2)_{-y}^2(\pi/2)_{-x}^2(\pi/2)_{y}^3-
\tau_2/2-(\pi)_x^{2,3}-\tau_2/2-(\pi/2)_{x}^3 \nonumber \\ &&(\pi/2)_{-x}^2(\pi/2)_{y}^2-
\tau_3/2-(\pi)_x^{1,2}-\tau_3/2-(\pi/2)_{-y}^2, \nonumber
\end{eqnarray}
where $\tau_1=1/(2J_{12})$, $\tau_2=1/(2J_{23})$ and $\tau_3=3/(2J_{12})$.

The parity gate was implemented using on the 3-qubit system of $^{13}$C labeled alanine. 
The three $^{13}$C nuclei of alanine 
 constitute a 3-qubit system.  The protons were decoupled by a standard decoupling sequence.
At 11.4 Tesla, the carbon-13 Larmor frequency differences are $\vert\nu_1-\nu_2\vert$= 15755 Hz, 
$\vert\nu_2-\nu_3\vert$=  4325 Hz and
$\vert\nu_1-\nu_3\vert$= 20080 Hz. 
The J-couplings of the system are $J_{12}=54$ Hz, $J_{13}=1.4$ Hz and $J_{23}=35.1$ Hz.
 The equilibrium spectrum of each carbon is given in figure 4(a).
 Starting from thermal equilibrium, the pulse sequence 
of 3(b) was applied. Finally, a gradient pulse was applied to destroy any unwanted coherence created by the 
imperfection of pulses. Since we started from thermal equilibrium state, the result is encoded in the 
final populations. 
Three separate experiments were performed with readout $(\pi/2)$ pulses on the 
three qubits respectively:(i) $(\pi/2)^1$, (ii) $(\pi/2)^2$ and (iii) $(\pi/2)^3$.
 The readout pulse applied to one qubit at a time, converts the differences in 
populations to amplitudes of the observable single quantum coherences. The spectra obtained in the three experiments 
is given in  figure 4(b). 
From the truth table it is evident that this gate interchanges the populations of
$\vert 010 \rangle \leftrightarrow \vert 011 \rangle$ and
$\vert 101 \rangle \leftrightarrow \vert 100 \rangle$.
The expected spectrum (corresponding to the final populations shown in figure 5(a)) is given below each spectrum 
in figure 4(c). This confirms the  implementation of the Parity gate, with maximum  error of 9$\%$.  

\section{Fanout gate}
 Fanout gate adds (addition modulo 2) a control bit onto 'n' target bits  (fig. 6(a)) \cite{ic}.
 The truth table of a fanout gate on a 3-qubit system where first qubit is the control qubit
and other two qubits are target, is given in table 3. It may be noted, that the (classical) value of
control bit is copied or 'fanned out' to the target bits if the target bits are initially in $\vert 0\rangle$ state
\cite{sha4}. 
 However, if the control bit is in coherent superposition, the fanout gate creates entangled states \cite{sha3,sha4}.  
Fanout gate is an important gate for creating shallow quantum circuits \cite{sha3,sha4}. A (n+1) fanout
 gate can be built out of (n+1) parity gate by applying Hadamard gates on both sides of the parity gate 
on the first n-qubits (fig. 6(a)).
 The pulse sequence was constructed by combining the pulses and evolutions of individual operation.
The combined and simplified pulse sequence (fig. 6(b)) is 
\begin{eqnarray}
&&(\pi/2)_{-y}^1-\tau_1/2-(\pi)_x^{1,2}-\tau_1/2-(\pi/2)_{-y}^2(\pi/2)_{x}^2-
\tau_2/4-(\pi)_x^{1}-\tau_2/4-(\pi)_x^{1,2,3}-\tau_2/4-(\pi)_x^{1}-\tau_2/4 \nonumber \\&&-(\pi/2)_{y}^2(\pi/2)_{-x}^2-
\tau_3/2-(\pi)_x^{1,2}-\tau_3/2-(\pi/2)_{-y}^1, \nonumber
\end{eqnarray}
where $\tau_1=1/(2J_{12})$, $\tau_2=1/(2J_{23})$ and $\tau_3=3/(2J_{12})$.

 Fanout gate was implemented on the 3-qubit system of 1 fluorine and 2 protons in the molecule 4-flouro-6-nitro benzofuran. 
At the magnetic field of 11.4 Tesla, the resonant frequency of proton is 500.13 MHz and that of fluorine is 470.59 MHz.
The frequency difference between the two protons is 250 Hz. We have taken the fluorine as the 1st qubit. 
The J-couplings are $J_{12}=3.84$ Hz, $J_{23}=8.01$ Hz and $J_{13}=-8.1 $ Hz. The equilibrium flourine and 
proton specta are given in figure 7(a).
Starting from equilibrium, the fanout gate was implemented with the pulse sequence of fig. 6(b). 
Finally, a gradient pulse was applied to kill any unwanted coherences created by the
imperfection of pulses. Three separate experiments were performed with readout $(\pi/2)$ pulses on the
three qubits respectively:(i) $(\pi/2)^1$, (ii) $(\pi/2)^2$ and (iii) $(\pi/2)^3$.
 The obtained spectra are given in figure 7(b). This gate interchanges the populations of $\vert100\rangle
\rightarrow \vert111\rangle$ and $\vert110\rangle\rightarrow \vert101\rangle$ (Fig. 5(b)). 
The expected spectrum, shown as stick diagram (figure 7(c)), confirms 
the implementation of Fanout gate, with the accuracy of 90$\%$.  

\section{conclusion}
 In conclusion, we have demonstrated the implementation of quantum shallow circuits in 3-qubit systems by NMR. 
The full potential of quantum shallow circuits can be exploited in higher qubit systems, 
and more important, in systems with same J-couplings between the qubits \cite{sha4}. 
In such systems the $U_n$ gate can be performed in parallel on all the required qubits and computation will 
be speed up. For example, when inversion-on-equality gate 
was implemented in a 3-qubit systems (section-II), the implementation was parallel and 
faster because two of the J-couplings were identical.
 Unlike other systems where qubits are identified and addressed in space, in NMR,
 the qubits are identified and addressed in frequency-space. Hence, it is mandatory that the 
spectrum be well resolved in NMR. When systems with same J-couplings are used, the transitions between the 
states overlap and resolution is lost, causing problems with identification of states. 
 Nevertheless, 
the present work validates the efficiency of quantum shallow circuits which will be a useful experience 
when such circuits are implemented in other physical systems where qubits are addressed in space, like 
solid-state devices \cite{kane}, quantum dots \cite{dots}, photons \cite{ic} and trapped-ions \cite{db,ic,ion}. 
\section{acknowledgment} 
Useful discussions with Stephen A. Fenner are gratefully acknowledged. 
The use of DRX-500 NMR spectrometer funded by the Department of
Science and Technology (DST), New Delhi, at the Sophisticated Instruments Facility, Indian Institute of Science, 
Bangalore, is gratefully acknowledged. AK acknowledges "DAE-BRNS" for the award of "Senior Scientists scheme", 
and DST for a research grant on "Quantum Computing using NMR techniques".

\pagebreak
\hspace*{4cm}
Table 1: Truth table of Inversion on equality gate\\\\
\hspace*{6cm}
\begin{tabular}{|c|c|} \hline
INPUT&OUTPUT \\ \hline
~~~~0~~~~0~~~~0~~~~&~~~~-0~~~~0~~~~0~~~~\\
~~~~0~~~~0~~~~1~~~~&~~~~0~~~~0~~~~1~~~~\\
~~~~0~~~~1~~~~0~~~~&~~~~0~~~~1~~~~0~~~~\\
~~~~0~~~~1~~~~1~~~~&~~~~0~~~~1~~~~1~~~~\\
~~~~1~~~~0~~~~0~~~~&~~~~1~~~~0~~~~0~~~~\\
~~~~1~~~~0~~~~1~~~~&~~~~1~~~~0~~~~1~~~~\\
~~~~1~~~~1~~~~0~~~~&~~~~1~~~~1~~~~0~~~~\\
~~~~1~~~~1~~~~1~~~~&~~~~-1~~~~1~~~~1~~~~\\
\hline
\end{tabular}
\pagebreak

\hspace*{5cm}
Table 2: Truth table of Parity gate\\\\
\hspace*{6cm}
\begin{tabular}{|c|c|} \hline
INPUT&OUTPUT \\ \hline
~~~~0~~~~0~~~~0~~~~&~~~~0~~~~0~~~~0~~~~\\
~~~~0~~~~0~~~~1~~~~&~~~~0~~~~0~~~~1~~~~\\
~~~~0~~~~1~~~~0~~~~&~~~~0~~~~1~~~~1~~~~\\
~~~~0~~~~1~~~~1~~~~&~~~~0~~~~1~~~~0~~~~\\
~~~~1~~~~0~~~~0~~~~&~~~~1~~~~0~~~~1~~~~\\
~~~~1~~~~0~~~~1~~~~&~~~~1~~~~0~~~~0~~~~\\
~~~~1~~~~1~~~~0~~~~&~~~~1~~~~1~~~~0~~~~\\
~~~~1~~~~1~~~~1~~~~&~~~~1~~~~1~~~~1~~~~\\
\hline
\end{tabular}
\pagebreak

\hspace*{5cm}
Table 3: Truth table of Fanout gate\\\\
\hspace*{5cm}
\begin{tabular}{|c|c|} \hline
INPUT&OUTPUT \\ \hline
~~~~0~~~~0~~~~0~~~~&~~~~0~~~~0~~~~0~~~~\\
~~~~0~~~~0~~~~1~~~~&~~~~0~~~~0~~~~1~~~~\\
~~~~0~~~~1~~~~0~~~~&~~~~0~~~~1~~~~0~~~~\\
~~~~0~~~~1~~~~1~~~~&~~~~0~~~~1~~~~1~~~~\\
~~~~1~~~~0~~~~0~~~~&~~~~1~~~~1~~~~1~~~~\\
~~~~1~~~~0~~~~1~~~~&~~~~1~~~~1~~~~0~~~~\\
~~~~1~~~~1~~~~0~~~~&~~~~1~~~~0~~~~1~~~~\\
~~~~1~~~~1~~~~1~~~~&~~~~1~~~~0~~~~0~~~~\\
\hline
\end{tabular}
\pagebreak

\hspace{6cm}\large{FIGURE CAPTIONS}
\\\\
FIG. 1: (a) The 3-qubit quantum circuit of inversion-on-equality gate `U$_3$'. U$_3$ gate flips the state 
in which all the qubits are in the same state. Hence in a 3-qubit case $\vert 000\rangle$ and $\vert 111\rangle$ 
will be flipped. (b) The complete quantum circuit. The pseudo-Hadamard pulses were 
applied to create coherent superposition of the qubits.
(c) Pulse sequence for implementation of (a). The thin and thick bars respectively represent $(\pi/2)$ 
and  $(\pi)$ pulses. The pulse sequence can be broken into two exclusive parts which is 
seperated by dashed lines. In the first part, the system evolves under all the J-couplings, $J_{12}$, $J_{13}$ 
and $J_{23}$ for $\tau_1=1/(2J_{12})=1/(2J_{13})$, while  in the second part, the system evolves 
under $J_{23}$-coupling for $\tau=(1/(2J_{23})-\tau_1)$. This completes a $(1/2J)$ evolution under all the J-couplings.\\\\

FIG. 2: (a) Equilibrium proton spectrum of 1-bromo, 2,3-dichloro benzene obtained after a hard $(\pi/2)_y$ pulse. 
(b) Ouput spectrum after implementation of the U$_3$ gate (pulse sequence of Fig. 1(c)).
(c) Stick diagram of the expected spectrum  corresponding to the density matrix of Eq.9 of the text.\\\\

FIG. 3: (a) The 3-qubit quantum circuit of Parity gate. A n-qubit parity gate adds (n-1) qubits to 
the n$^{th}$ qubit. Starting from a state $\vert x_1x_2x_3\rangle$, the Hadamard gate on the 
second qubit prepares it in the state ($\vert 0\rangle+ (-1)^{x_2}\vert 1\rangle$). After $U_2$, the 
state of the second qubit is in superposition (y or -y axis of Bloch sphere) 
($i^{x_1}\vert 0\rangle+ i^{1-x_1}(-1)^{x_2}\vert 1\rangle$ and it is brought to the 
z-axis of the Bloch sphere by $HS$ gates. When the CNOT qubit is applied, the control (2nd) qubit is in the state 
$i^{x_1}\vert x_1\oplus x_2 \rangle$ and hence after CNOT gate the target (3rd) qubit is in the state 
$i^{x_1}\vert x_1\oplus x_2 \oplus x_3\rangle$. The rest of the circuit is then needed to cancel the phase factor $i^p$.   
(b) Pulse sequence for implementation of (a). 
Pseudo-Hadamard gate is implemented by a $(\pi/2)_{-y}$ pulse. The phase gates are implemented by 
$(\pi/2)_{-y}(\pi/2)_{x}(\pi/2)_{y}$. The CNOT gate between 2nd and 3rd qubit can be implemented by
$(\pi/2)_y^3-\tau/2-(\pi)_x^{2,3}-\tau/2-(\pi/2)_x^3$. These pulses are combined to give the above pulse sequence.\\\\

FIG. 4: (a) Equilibrium $^{13}$C-spectra (proton decoupled) of the 3-qubit system of $^{13}$C-labeled alanine. 
(b) The spectra after application of Parity gate followed by qubit-selective read out pulses. The intensities of 
$x_1$ and $x_2$ spectra have been scaled down by a factor of 2, and to compare these intensities 
with the equlibrium spectra of fig. 4(a), should be multiplied by 2. 
(c) Expected spectra from  the final populations given in Fig. 5(a).\\\\

FIG. 5: (a)Energy level diagram and the initial (equilibrium) and final (after implementing
parity gate) populations of the states in a 3-qubit system.
 There is a exchange of populations between the states
$\vert010\rangle \leftrightarrow \vert011\rangle$ and $\vert100\rangle \leftrightarrow \vert101\rangle$
The final populations of each level are given in the bracket along with the initial populations outside the brackets.
(b) Equilibrium and final populations after implementation of fanout gate.  There is an exchange of populations between the states
$\vert001\rangle \leftrightarrow \vert011\rangle$ and $\vert100\rangle \leftrightarrow \vert101\rangle$ by this gate.\\\\

FIG. 6: (a) The 3-qubit quantum circuit of fanout gate.  A 3-qubit fanout
 gate can be built out of 3-qubit parity gate by applying pseudo-Hadamard gates on both sides of the parity gate
on the first 2 qubits. Hence, starting with the circuit of figure 3(a), applying Hadamard gates on both sides and
after some simplification we get the above circuit.
(b) Pulse sequence for implementation of (a).
Pseudo-Hadamard gate is implemented by a $(\pi/2)_{-y}$ pulse. The phase gates are implemented by
$(\pi/2)_{-y}(\pi/2)_{x}(\pi/2)_{y}$. The CNOT gate between 2nd and 3rd qubit can be implemented by 
$(\pi/2)_{y}^2-
\tau_2/4-(\pi)_x^{1}-\tau_2/4-(\pi)_x^{1,2,3}-\tau_2/4-(\pi)_x^{1}-\tau_2/4-(\pi/2)_{x}^2$. 
These pulses are combined to give the above pulse sequence.\\\\

FIG. 7: (a) Equilibrium flourine and proton spectra of 4-fluoro-6-nitro benzofuran.
The peak is shown by (*) belongs to the solvent. 
(b) The spectra after application of fanout gate obtained by qubit selective read out pulses.
(c) Expected spectra from the final populations given in Fig. 5(b).\\\\

\pagebreak
\begin{figure}
\epsfig{file=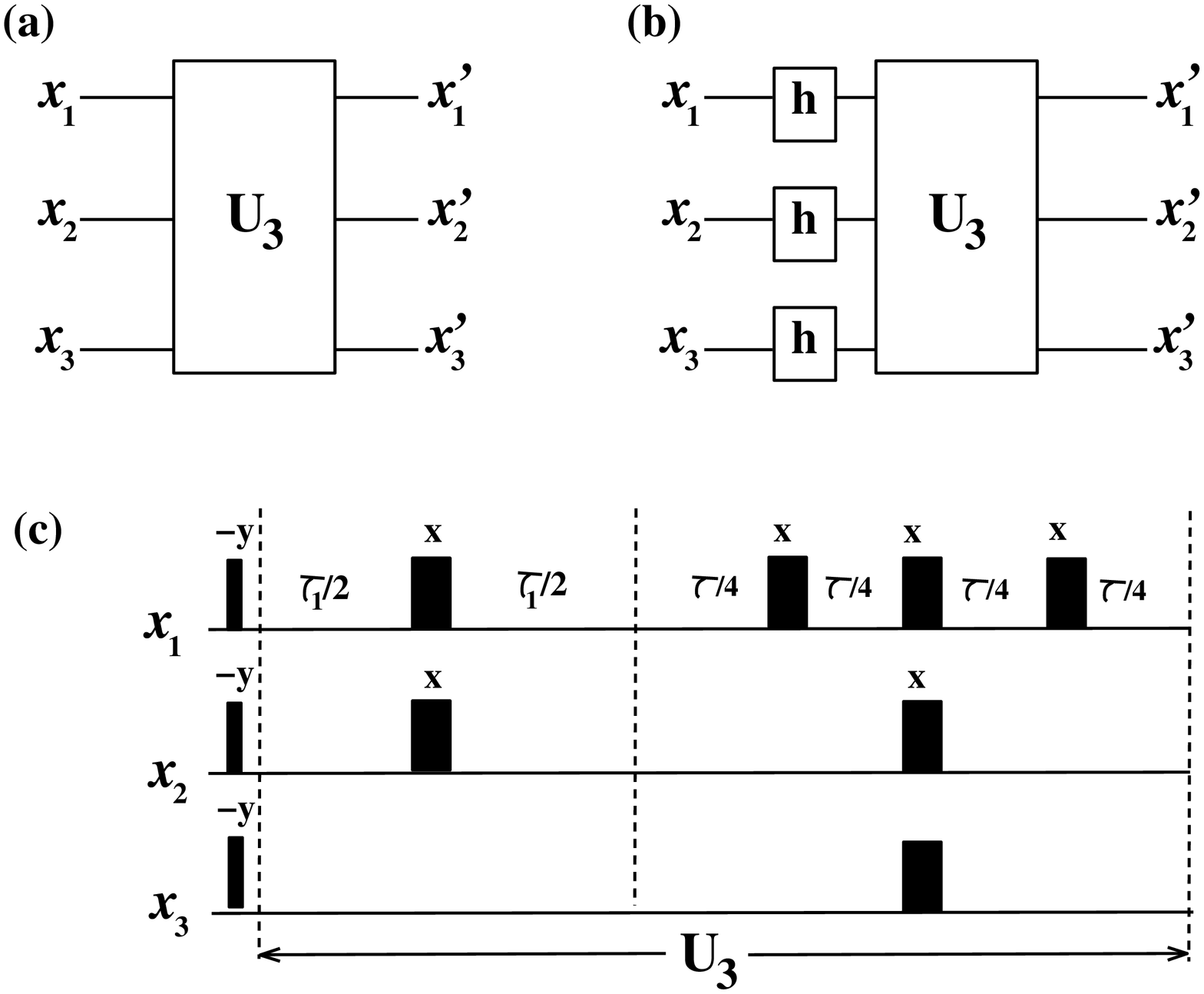,height=16cm,angle=270}
\end{figure}
\vspace{2cm}
\hspace{6cm}
{\huge Figure 1}
\pagebreak
\begin{figure}
\epsfig{file=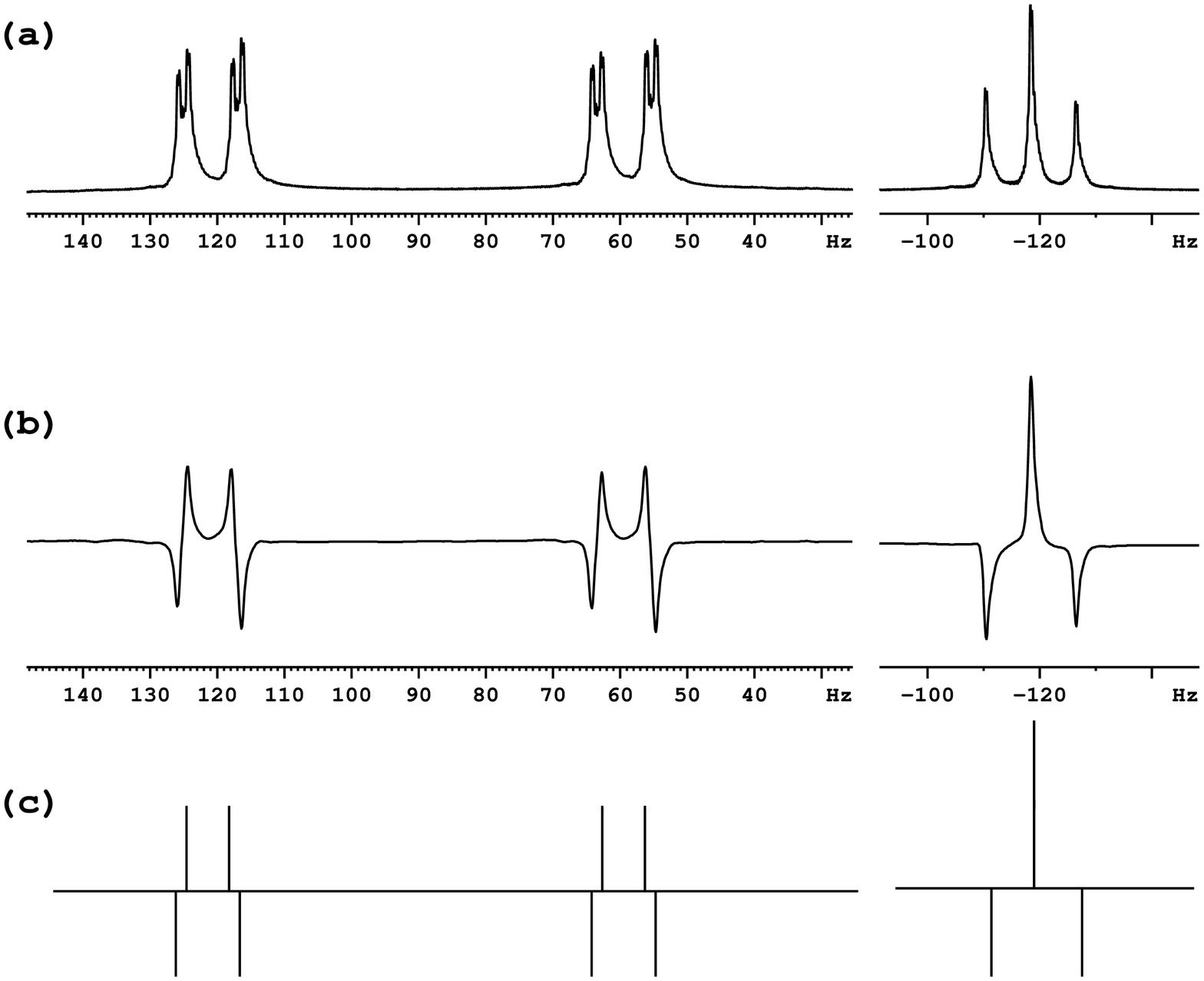,height=16cm,angle=270}
\end{figure}
\vspace{2cm}
\hspace{6cm}
{\huge Figure 2}
\pagebreak
\begin{figure}
\epsfig{file=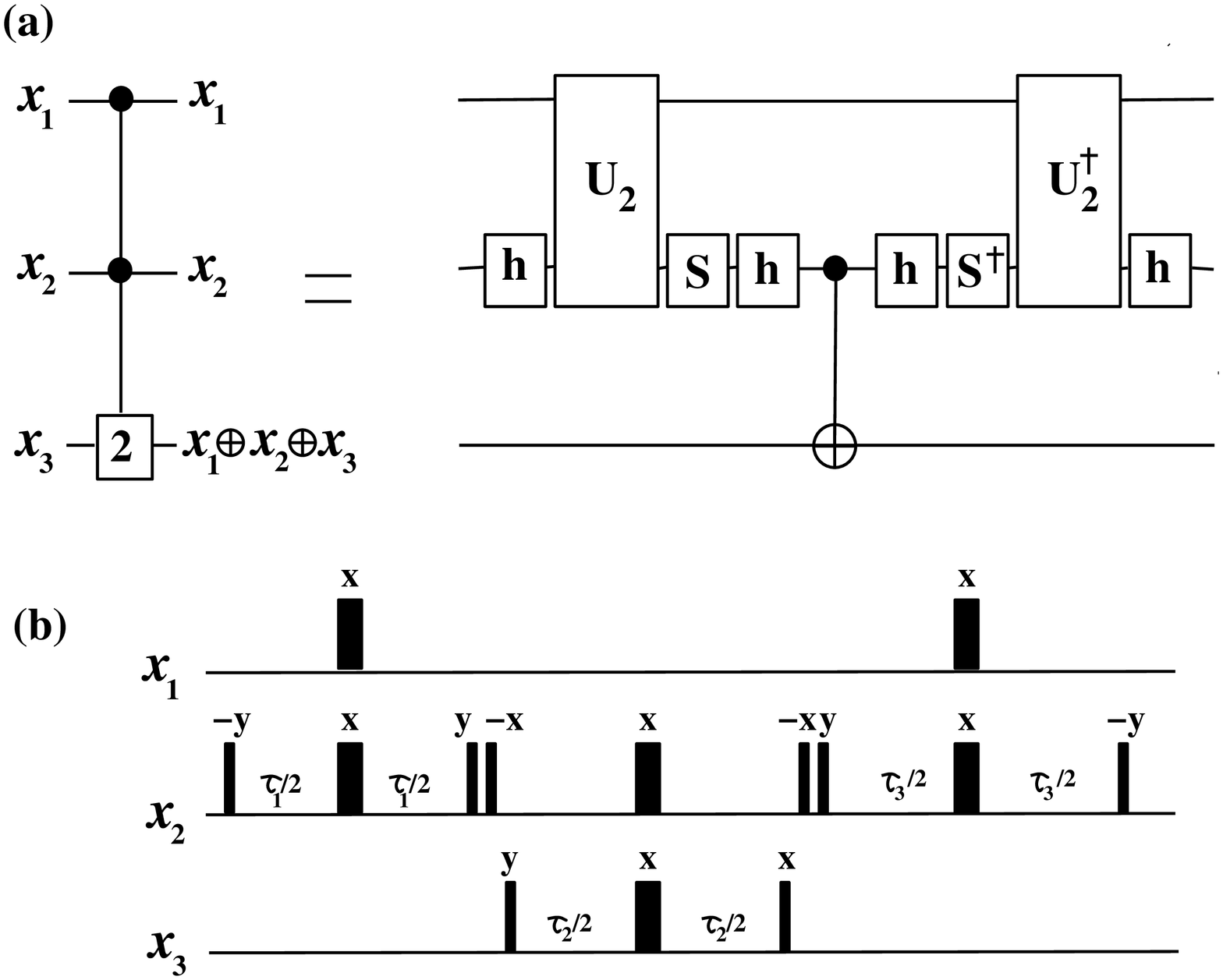,height=16cm,angle=270}
\end{figure}
\vspace{2cm}
\hspace{6cm}
{\huge Figure 3}
\pagebreak
\begin{figure}
\epsfig{file=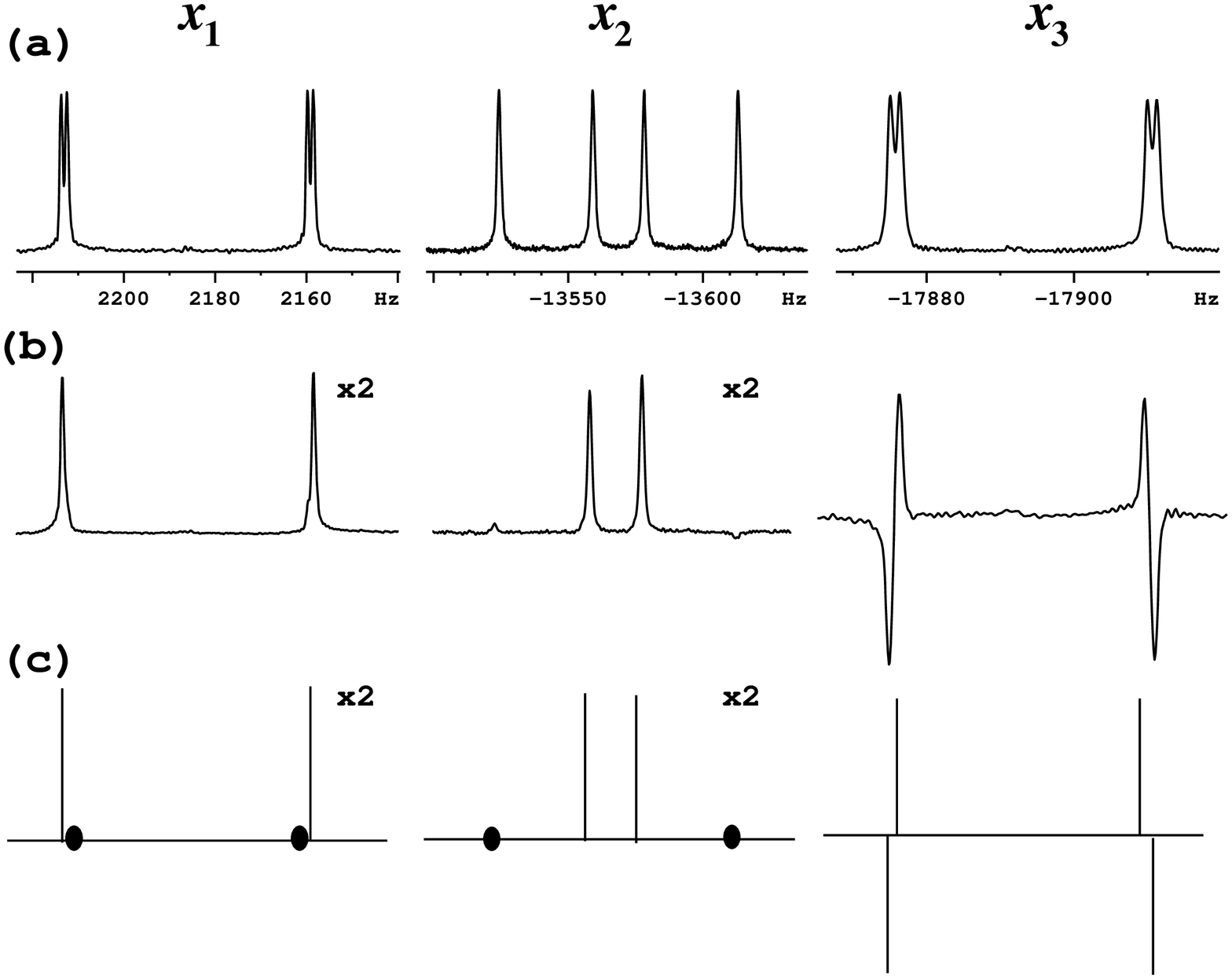,height=16cm,angle=270}
\end{figure}
\vspace{2cm}
\hspace{6cm}
{\huge Figure 4}
\pagebreak
\begin{figure}
\epsfig{file=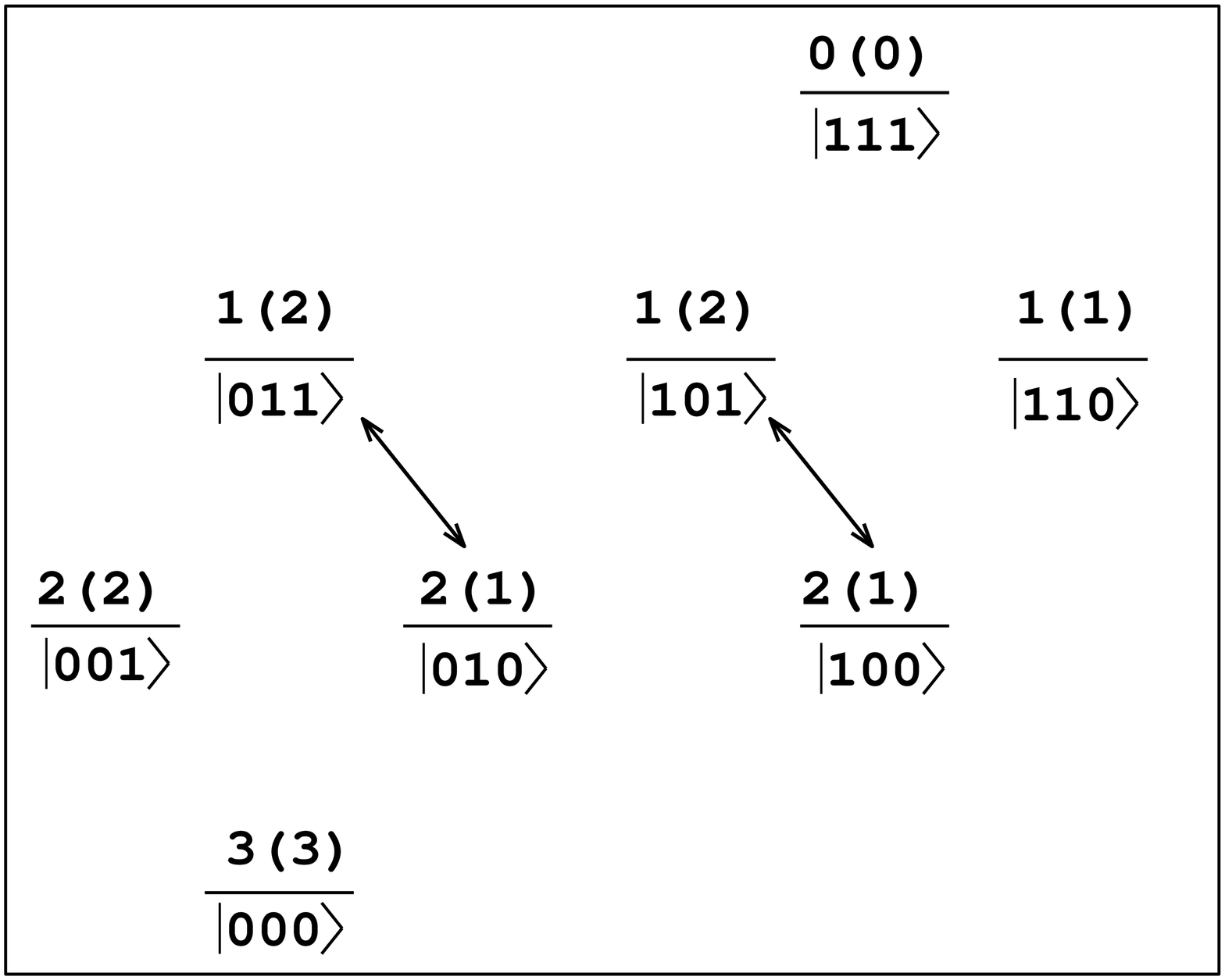,height=8cm,angle=270}
\end{figure}
\hspace{3cm}
{\huge (a)}
\begin{figure}
\epsfig{file=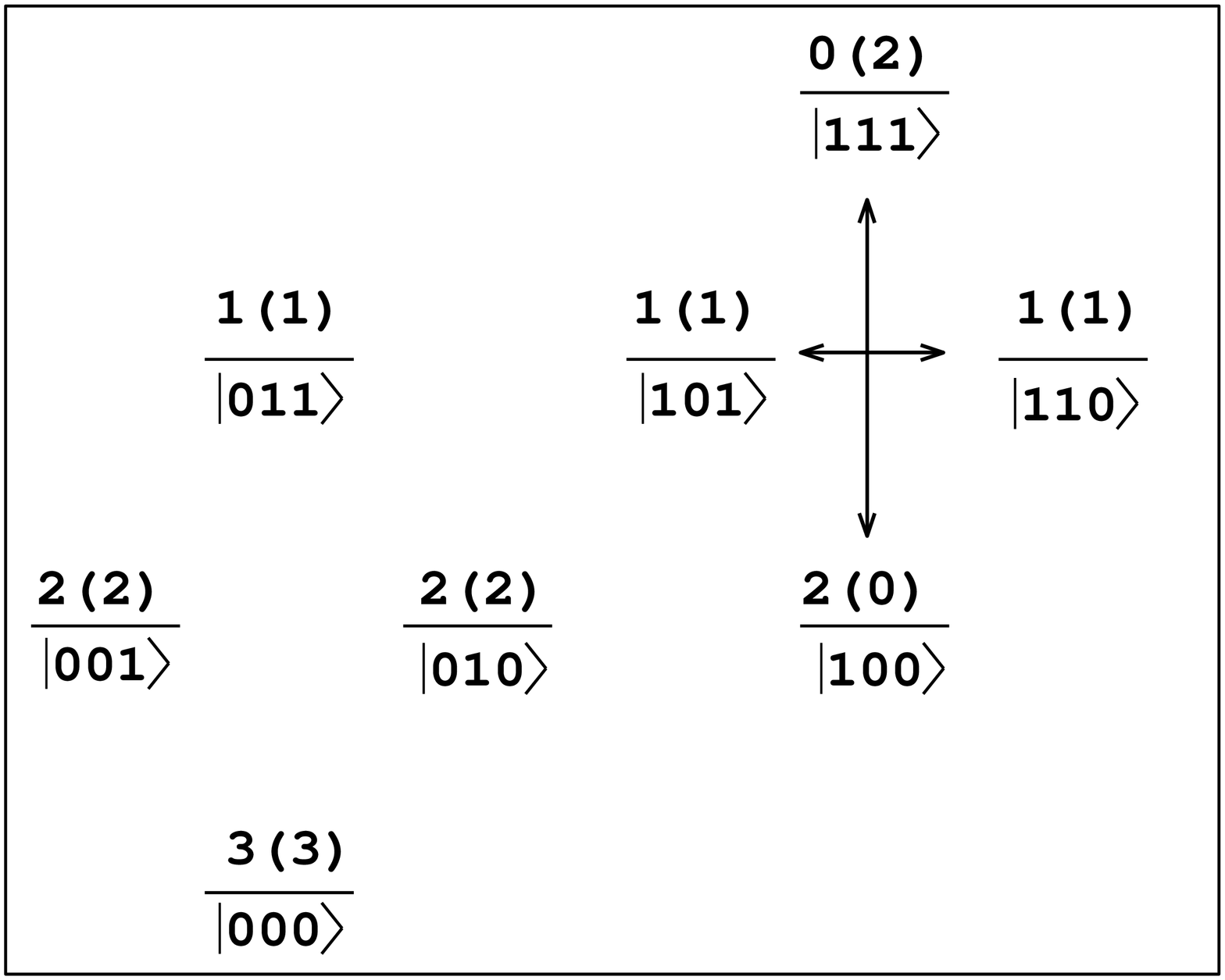,height=8cm,angle=270}
\end{figure}
\hspace{3cm}
{\huge (b)}\\\\
\hspace*{6cm}
{\huge Figure 5}
\pagebreak
\begin{figure}
\epsfig{file=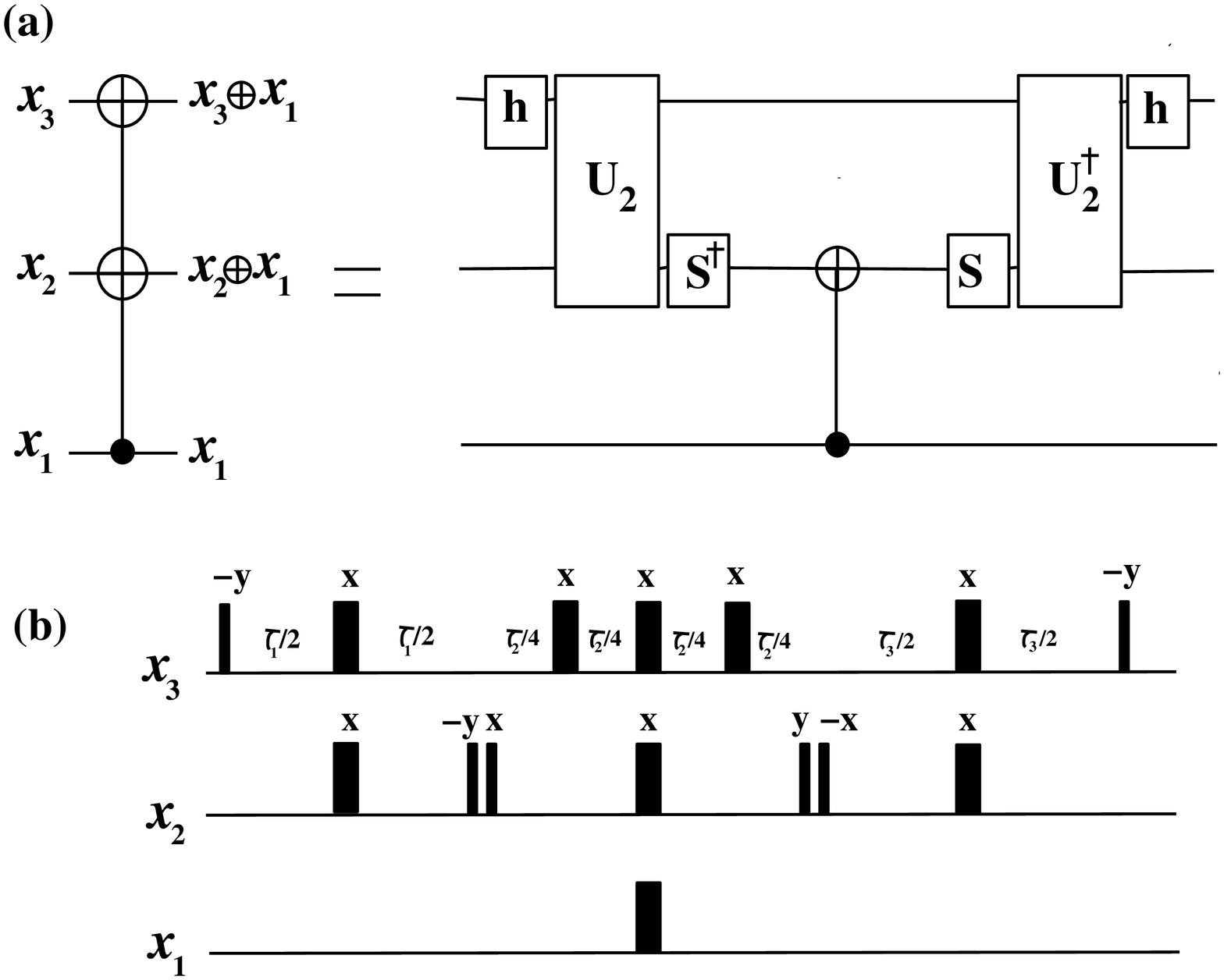,height=16cm,angle=270}
\end{figure}
\vspace{2cm}
\hspace{6cm}
{\huge Figure 6}
\pagebreak
\begin{figure}
\epsfig{file=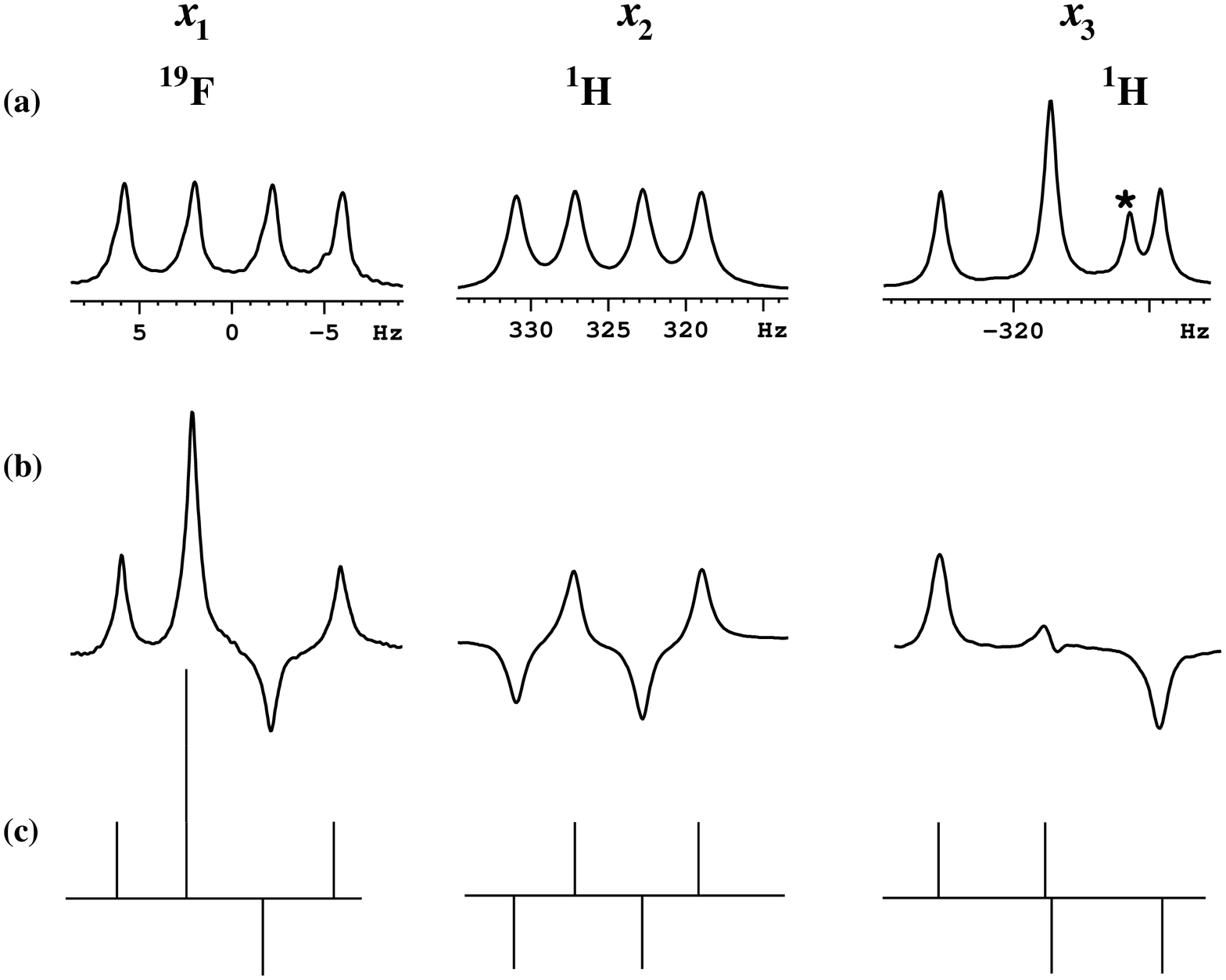,height=16cm,angle=270}
\end{figure}
\vspace{2cm}
\hspace{6cm}
{\huge Figure 7}

\end{document}